\documentclass[12pt]{article}
\usepackage[dvips]{graphicx}
\usepackage{amsmath}
\usepackage{amssymb}
\usepackage{amstext}
\usepackage{amsfonts}
\usepackage{slashbox}
\usepackage{indentfirst}
\begin{document}

\title{Bose-Einstein condensation in strong-coupling quark color superconductor near flavor $SU(3)$ limit\footnote{This work is supported by the National Natural Science Foundation of China ( NSFC ) under Grant No.10875058.}}

\author{ Xiao-Bing Zhang $^{1}$ , Chun-Fu Ren $^{1}$\footnote{Corresponding author, e-mail: nkupc2002@mail.nankai.edu.cn.} and Yi Zhang $^{2}$ \\
{\small $^{1}$ School of Physics, Nankai University, Tianjin  300071, China}\\
{\small $^{2}$ Department of Physics, Shanghai Normal University, Shanghai 200230, China}}

\maketitle

\begin{abstract}

\rm \noindent Near the flavor SU(3) limit, we propose an analytical description for color-flavor-locked-type Bardeen-Cooper-Schrieffer (BCS) phase in the Nambu Jona-Lasinio (NJL) model. The diquark behaviors in light-flavor and strange-flavor-involved channels and
Bose-Einstein condensation (BEC) of bound diquark states are studied.
When the attractive interaction between quarks is strong enough,
a BCS-BEC crossover is predicted in the environment with color-flavor-locked pairing pattern.
The resulting Bose-Einstein condensed phase is found to be an intergrade phase before the emergence of
the previous-predicted BEC phase in two-flavor quark superconductor.
\end{abstract}


\baselineskip 12pt

\section{\bf Introduction}

Dense quark matter has been an exciting topic since quark color superconducting phases could form the ground states
of quantum chromodynamics (QCD). At asymptotically high baryon
density ( equivalently, the quark chemical potential $\mu$ is
extremely large ) and low temperature, the ground state of
three-flavor matter is believed to correspond to the
color-flavor locked (CFL) phase \cite{alf}.
The appearance of CFL is completely due to the
Bardeen-Cooper-Schrieffer (BCS) mechanism, i. e. the weak attraction
between quarks leads to that quarks participate in the pairing on
the Fermi surface. In the CFL phase, the original QCD symmetry
$SU(3)_{color}\times SU(3)_{L} \times SU(3)_{R}$ is broken down to
the diagonal subgroup $SU(3)_{color+L+R}$ \cite{alf}, similar as the
chiral breaking pattern in the low-energy QCD. As the consequence of this symmetry breaking pattern, there exist the
Goldstone-type modes in this phase, similar as the spectrum of the lowest-lying pseudo-Goldstone meson in hadronic matter.
Based on these similarities mainly, a conjecture of the continuity between hadronic matter and quark matter was supposed by Sch\"{a}fer and F. Wilczek
\cite{sw99}. Until now, a fully understanding of the QCD phase diagram at finite $\mu$ ( including the hadron-quark transition ) is immature.
In recent years, the detailed phase structure of quark color superconductivity has been widely studied ( see, e. g. Refs.\cite{alf03},\cite{bubar} and \cite{asrs07} for reviews ).

Generally, candidates of the second-densest ground state of QCD include
the less-symmetric BCS phases as well as the non-BCS phases ( like a crystalline superconducting phase ).
They might become energetically more favorable than CFL in the densities relevant for compact star phenomenology.
Qualitatively, emergence of a less-symmetric BCS phases could be
understood as a phenomenon near the Fermi surfaces of quarks.
In the typical BCS region, effects of the current quark masses, charge neutrality and $\beta$-equilibrium
turn on mismatches between
the Fermi momenta for different quark species, which makes the less-symmetric BCS phases possible.
In the situation with not-very-large $\mu$, in particular, the explicitly
chiral breaking needs to be taken into account. Ignoring the current quark masses for light flavors ( $u$ and
$d$ ), the effect of strange-flavor current mass $m_s$ can be attributed
to the effective chemical potential for strange-flavor quarks.
When $m_s \ll \mu$ is satisfied, it is usually expressed as
$\mu^s_{eff} = \mu-\frac{m_s^2}{2\mu}$ at the leading order \cite{alfj02}.
The mismatch ${m_s^2}/{2\mu}$ plays a key role in formation of color-flavor unlocking and emergence of a two-flavor color superconducting phase ( 2SC )\cite{alf,alfj02,buba02,rust05}.

On the other hand, the strong-coupling nature of QCD is manifested gradually in intermediate density region.
Assuming that color superconducting matter still exists, quark Cooper pairs ( diquarks ) should be regarded as the key degrees of freedom.
With increasing strength of attractive quark-quark interaction and/or decreasing quark chemical potential, there might be tightly bound states consisting of diquarks. Naturally a theoretical possibility arises that such kind of bosonic bound states undergo the
Bose-Einstein condensation (BEC). In condensed matter physics, it has been noticed that BCS of Cooper
pairs and BEC of bound difermion molecules are two sides of the same coin.
BEC displays a strong correlation in the coordinate space whereas BCS does a strong correlation in the momentum space.
The BCS-BEC crossover physics in the scope of quark color superconductivity has attracted much attention and been investigated by a variety of QCD models
\cite{mats00,abuki02,itak03,abuki05,baym06,zhuang07,wang07,abuki07,kita08,abuki10}.
For the strong-coupling regime, people usually apply the phenomenological models such as the Nambu Jona-Lasinio (NJL) model with both quark-antiquark and quark-quark interactions.
For instance, the crossover physics was studied in such a three-flavor NJL model \cite{kita08}. As the result, the BEC occurrence was predicted for low $\mu$ in the situation with the realistic current masses ( i. e. $m_s=120$MeV and $m_{u,d}=5$MeV ).
By incorporating the axial anomaly ( as an interplay between the quark-antiquark and diquark condensates ), it has been argued more recently that the BEC occurrence might be enhanced \cite{abuki10}. There, it was found that BEC may be realized at relatively large $\mu$ in the situation with equal masses $m_u=m_d=m_s$.

In the present work, we will focus on three-flavor superconducting matter with small $m_s$ and vanishing $m_{u,d}$.
By regarding the SU(3) flavor symmetry as approximately correct, as the starting point, it is required that $m_s$ is far smaller than the constituent quark masses obtained in an NJL model. For the CFL phase, it is safe to assume that there exist the common constituent quark mass $\overline{M}$ ( defined below ) for all quark species. Correspondingly, the dispersion laws for diquarks take the similar expressions as those given in an ideal CFL phase defined at extremely large $\mu$. This point allows us to do analytical calculations for three-flavor quark superconductors.
Meanwhile, an explicit chiral breaking reflected by $m_s \neq 0$
needs to be incorporated in the analytical way. For the purpose of exploring the BCS-BEC crossover physics, we concern the quark chemical potential which has the same order as the constituent mass $\overline{M}$.
Because $m_s \ll \overline{M}$ has been presumed, the condition $m_s
\ll \mu $ is satisfied throughout the present work. This point allows us to extrapolate the treatment
which was widely used in the typical BCS region ( with large $\mu$ ) to the possible BEC region ( with $\mu \sim \overline{M}$ ).

Along the above-mentioned lines, exploring the BEC occurrence and its response to the strange mass $m_s$ become available technically. Keeping it in mind that
three-flavor system is considered near the flavor $SU(3)$ limit, we construct
an analytical description for the CFL-type BCS superconducting phase in section 2.
In section 3, the behaviors for two kinds of paired quarks ( the paired quarks made up of
light- and strange-flavor quark species ) are investigated at the mean-field level.
By taking color-flavor unlocking transition into account, we observe that there exists a BCS-BEC crossover before the CFL pairing pattern becomes broken completely. The resulting BEC phase is different from the previous-predicted one in two-flavor superconductor.
Its existence is discussed to be reasonable as long as the coupling strength of diquark interaction is strong enough and the flavor
$SU(3)$ symmetry holds approximately.

\section{\bf Formalisms}
\subsection{ Ideal CFL }

Our starting point is an ideal CFL phase in the flavor $SU(3)$  limit ( the chiral limit ), which exists actually at for extremely high densities.
With vanishing current quark masses,
all paired quark species are degenerate so that one obtain the common constituent quark mass $M$ and the common color-superconducting gap $\Delta$. In this phase, the dispersion laws for diquarks are given in the forms of
\begin{eqnarray}
 E^{\pm}(\Delta)\equiv E^{\pm}(M,\mu,\Delta) = \sqrt{(\sqrt{p^2+M^2}\mp\mu)^2 + \Delta^2},\label{e+-}
 \end{eqnarray}
and $E^{\pm}(2\Delta)\equiv E^{\pm}(M,\mu,2\Delta)$. For relativistic system of quarks,
$\sqrt{p^2+M^2}$ denotes the single-particle energy while $E^{+}$ and $E^{-}$ correspond to the energies of quasiparticle excitations for quark and antiquark, respectively.
Within the framework of the NJL-type models, it is well known that
the thermodynamic potential is
\begin{eqnarray}
\frac{3 M^2}{8 G_s}+ \frac{3\Delta^2}{4 G_d} - 8\int
\frac{d^3p}{(2\pi)^3}\left[ E^+(\Delta) + E^-(\Delta)\right] - \int
\frac{d^3p}{(2\pi)^3}\left[ E^+(2\Delta) + E^-(2\Delta)\right] \, , \label{icfl}
\end{eqnarray}
at zero temperature, where $G_s$ and $G_d$ are the quark-antiquark
and diquark coupling constants respectively. By minimizing the
thermodynamic potential, the gap equations for the chiral and color
condensates are obtained and their solutions correspond to the
constituent quark mass $M$ and the color superconducting gap
$\Delta$.

In general the paired quarks need to be differentiated and
the CFL pairing ansatz reads \cite{alf}
\begin{eqnarray}
{\langle q^\alpha_i C\gamma_5 q^\beta_j\rangle} \sim \Delta_1
\epsilon^{\alpha\beta 1}\epsilon_{ij1} + \Delta_{2}
\epsilon^{\alpha\beta 2}\epsilon_{ij2} +
\Delta_{3}\epsilon^{\alpha\beta 3}\epsilon_{ij3} ,  \label{ansatz}\end{eqnarray}
where $(\alpha,\beta)$ stands for the color indices $(r,g,b)$ and $(i,j)$ for the flavor indices $(u,d,s)$.
In the situation with $m_{u,d}=0$ and $m_s \neq 0$, there exist the two species of paired quarks
related with the color-superconducting gaps
$\Delta_{3}$ and $\Delta_{1,2}$ respectively.
Equivalently, there exist the two kinds of diquarks which behave as the
bosons composed of paired fermions.
The channel characterized with $\Delta_{3}$ describes the pairings
between the light-flavor ( and the non-blue-color ) quarks. In the ideal CFL phase, $\Delta_{3}$ is equal to the common value $\Delta$ so that the thermodynamic-potentialcontribution from such kind of diquarks is written as
\begin{eqnarray}
\Omega_3 (\Delta) \equiv \Omega_3 (M,\mu,\Delta) = -4\int
\frac{d^3p}{(2\pi)^3}\left[E^+(\Delta) + E^-(\Delta)\right].\label{o3}
\end{eqnarray}
The $\Delta_{1,2}$ channel usually describes the pairings
between the light- and strange-flavor quarks. According to
Eq.(\ref{ansatz}), it involves both the pattern where two quarks paired each
other and that where three quarks paired mutually ($ru$, $gd$ and
$bs$). Correspondingly, the thermodynamic potential contribution is
\begin{eqnarray}
\Omega_{1,2}(\Delta)&\equiv&\Omega_{1,2} (M,\mu,\Delta)\nonumber\\&=&-4\int
\frac{d^3p}{(2\pi)^3}\left[E^+(\Delta) + E^-(\Delta)\right]-\int
\frac{d^3p}{(2\pi)^3}\left[E^+(2\Delta) + E^-(2\Delta)\right] \label{o12}
\end{eqnarray}
Eqs.(\ref{o3}) and (\ref{o12}) as well as the mean-field
terms involving $M$ and $\Delta$ consist of the total thermodynamic
potential Eq.(\ref{icfl}) for the ideal CFL phase.
Besides, the number density of the paired quarks can be obtained by
taking the partial derivative of Eq.(\ref{icfl}) with respect to
$\mu$, namely
\begin{eqnarray} \rho=-\frac{\partial \Omega_{1,2}(\Delta)}{\partial
\mu}-\frac{\partial \Omega_{3}(\Delta)}{\partial \mu}.
 \label{rho}
\end{eqnarray}
There $-\partial\Omega_{1,2}(\Delta)/\partial\mu$ represents
the density of the quark participating in the pairing with
$\Delta_{1,2}$, while $-\partial\Omega_{3}(\Delta)/\partial\mu$ does
that in the $\Delta_{3}$ pairing.

\subsection{CFL near the flavor $SU(3)$  limit}

In principle, the unequal current masses give different chiral
condensates and different constituent masses at mean-field level.
Within the framework of NJL-type models, it leads to that
the dispersion laws Eq.(\ref{e+-}) are not valid and
the thermodynamic potential for CFL becomes too complicated to be calculated
analytically \cite{buba02}. To avoid this complication, we will consider the
CFL phase near the flavor $SU(3)$  limit,
where $m_{s}$ is small enough with respect to the constituent quark masses.
In this case, it is safe to assume that there exist the approximately same mean-field values of chiral condensate for three flavors.
To distinguish from $M$ defined in the ideal CFL phase, the common value is regarded as
the average quark mass $\overline{M}$.
At the same time, effect of a small $m_{s}$ needs to be considered by taking into account
difference of effective chemical potentials for the light and strange flavors .
It is noted that the average values of effective chemical potentials, i. e. the
chemical potentials for paired quarks, are relevant for studying the diquark behaviors.
Without losing generality, we introduce the average chemical potentials
$\overline{\mu}_{1,2}$ and $\overline{\mu}_{3}$ for the paired quarks
in the $\Delta_{1,2}$ and $\Delta_{3}$ channels respectively.
Equivalently, the chemical potentials for the two kinds of diquarks read $2\overline{\mu}_{1,2}$
and $2\overline{\mu}_{3}$ in the CFL phase.

With the help of $\overline{M}$ and $\overline{\mu}$, it is not necessary to require that
the formalisms of dispersion laws Eq.(\ref{e+-}) are changed
( since the common constituent mass has been assumed ).
For the light-flavor channel related with the gap $\Delta_{3}$,
the chiral-limit result $E^{\pm}(\Delta)$
should be replaced by
\begin{eqnarray}
E^{\pm}(\Delta_{3})\equiv
E^{\pm}(\overline{M},\overline{\mu}_{3},\Delta_{3})
 = \sqrt{(\sqrt{p^2+{\overline{M}}^2}\pm{\overline{\mu}_{3}})^2 + \Delta_{3}^2}.\label{e+-3}
\end{eqnarray}
In the $\Delta_{1,2}$ channel, we have
\begin{eqnarray} E^{\pm}(\Delta_{1,2})\equiv
E^{\pm}(\overline{M},\overline{\mu}_{1,2},\Delta_{1,2})
,\label{e+-12}
\end{eqnarray}
and $E^{\pm}(2\Delta_{1,2})\equiv E^{\pm}(\overline{M},\overline{\mu}_{1,2},2\Delta_{1,2})$.
Likewise, the thermodynamic-potential contributions from
the two kinds of diquarks can be rewritten as
$\Omega_{1,2}\equiv \Omega_{1,2}(\overline{M},
\overline{\mu}_{1,2},\Delta_{1,2})$ and $ \Omega_{3}\equiv
\Omega_{3}(\overline{M}, \overline{\mu}_{3},\Delta_{3} )$
respectively.

Instead of incorporating the flavor asymmetry into the matrix consisting of quark masses, we
concern the main influence of a small $m_s$ on the formal description of CFL.
In the typical BCS region, it is well known that flavor asymmetry can be
attributed to the mismatches between effective chemical potentials for different flavors. When
$m_s \ll \mu$, a small $m_s$ ( exactly $m_s^2/2\mu$ ) is
associated with the strangeness effective chemical potential. Physically, its introduction enhances the
strangeness content and reduces the number density of the paired
quarks involving strange flavor.
Provided that the changed density induced by $m_s^2/2\mu$ is $\delta\rho$, the density for the paired
quarks in the $\Delta_{1,2}$ channel should be decreased. Formally, it can be expressed by
\begin{eqnarray} \rho_{1,2}=-\frac{\partial \Omega_{1,2}}{\partial
\overline{\mu}_{1,2}}-\delta\rho , \label{rho12}
\end{eqnarray}
rather than $-\partial\Omega_{1,2}(\Delta)/\partial\mu$ given
in Eq.(\ref{rho}).
At the given $\mu$, it is reasonable to
assume that the total quark density is conserved.
In the CFL phase where all the quark species participate in the BCS pairing, $\delta\rho$ corresponds to the number-density transfer
between the two of pairing channels. In this sense,
the density of light-flavor paired quarks ( in the $\Delta_{3}$ channel ) becomes
\begin{eqnarray}
\rho_{3}=-\frac{\partial \Omega_{3}}{\partial
\overline{\mu}_{3}}+\delta\rho , \label{rho3}
\end{eqnarray}
correspondingly.

Before we proceed, several points are worth being notified. First,
we have ignored the effects of electric/color neutrality of quark matter although they might be connected with
the flavor asymmetry in the CFL phase ( see, e. g. \cite{kry03} ).
The gapless phenomenon is ignored for simplicity although it was found to be triggered by
$m_s \neq 0$ mainly \cite{alf04,zhang07}.
Secondly, our concerned phase is not the conventional CFL phase, but the CFL-type superconducting phase
near the flavor $SU(3)$ limit ( the chiral limit ).
As mentioned in Introduction, the condition $m_s \ll \mu$ ( exactly, $m_s \ll \overline{M}$ and $\overline{M} \sim \mu$ ) is satisfied formally and the effect of $m_s^2/2\mu$ is manifested in a similar way.
Moreover, it is important to
notice that the BCS-BEC crossover is not a phase transition in the proper sense. In vicinity of the crossover point, there exists
a smooth behavior in the physical quantities like the number densities. Therefore, the physical
influences in the BCS region like Eqs.(\ref{rho12}) and (\ref{rho3}) can be extrapolated to the possible BEC region.

Keeping it in mind that $\overline{\mu}_{1,2}$ and $\overline{\mu}_{3}$
correspond to the conjugate variables to
the densities $\rho_{1,2}$ and $\rho_{3}$ respectively,
we transform the physical influences into the expressions of average chemical potentials.
In the presence of $\mu$, the grand-canonical potential could be generally written as
\begin{eqnarray}
-\frac{1}{\beta} Tr (e^{-{\beta} (\hat{H}-\mu \hat{\rho})}) ,
\label{omega}\end{eqnarray}
where $\hat{H}$ denotes the
Hamiltonian operator and $\hat{\rho}$ does the particle number
operator. In the NJL-type models with the quark-antiquark and diquark interactions, one usually employs the Nambu-Gorkov propagator of quarks to derive
the thermodynamic potential for the color superconducting phase.
In the ideal CFL phase,
it has been shown that the quark-antiquark interacting term ( chiral condensate ) and
the external-source term ( which is related with $\mu$ ) enter the
diagonal components of the inverse propagator, while the diquark interacting term ( color
condensate ) enters the off-diagonal components \cite{buba02}.
This is why the dispersion relations of paired quarks and the CFL thermodynamic potential can be obtained analytically
in the flavor $SU(3)$  limit.
Now we turn to the situation near the $SU(3)$ limit. Since the common value of chiral condensate has been assumed,
there should not exist difference in the quark-antiquark interacting terms for three flavors.
Also, there is not the mixing between the diquark interacting terms
and the external-source term.
Therefore, introducing the two densities $\rho_{1,2}$ and $\rho_{3}$ does not lead to complications in the matrix structure of the
inverse propagator and the resulting dispersion relations. In this sense, we replace the term $\mu \hat{\rho}$ in Eq.(\ref{omega})
by the term $\overline{\mu}_{1,2}
\hat{\rho_{1,2}} +\overline{\mu}_{3} \hat{\rho_{3}}$, namely
\begin{eqnarray}
\mu {\rho}=\overline{\mu}_{1,2} {\rho_{1,2}} +\overline{\mu}_{3}
{\rho_{3}}, \label{murho}
\end{eqnarray}
at the mean-field level.
If the two species of paired quarks are
independent,
Eq.(\ref{murho}) may be simplified as
\begin{eqnarray}
-\mu \frac{\partial \Omega_{1,2}(\Delta)}{\partial \mu}=
\overline{\mu}_{1,2} {\rho_{1,2}},\label{murho12}
\end{eqnarray}
and
\begin{eqnarray} -\mu \frac{\partial
\Omega_{3}(\Delta)}{\partial \mu}= \overline{\mu}_{3} {\rho_{3}}.
\label{murho3}
\end{eqnarray}

In analogy with the ideal case, the CFL thermodynamic potential is
written as
\begin{eqnarray}
\Omega_{CFL}&=& \frac{3 \overline{M}^2}{8 G_s}+
\frac{2\Delta_{1,2}^2}{4 G_d} + \frac{\Delta_{3}^2}{4 G_d} +
\Omega_{1,2}(\overline{M}, \overline{\mu}_{1,2},\Delta_{1,2}) +
\Omega_{3}(\overline{M}, \overline{\mu}_{3},\Delta_{3} ) \, ,
\label{cfl}
\end{eqnarray}
near the flavor $SU(3)$  limit, which returns to Eq.(\ref{icfl}) when $\delta\rho$ vanishes. The
common quark masses $\overline{M}$ is deviated from $M$ (which is obtained in the ideal phase)
and needs to be recalculated by minimizing
Eq.(\ref{cfl}). Similarly,
the gaps $\Delta_{1,2}$ and $\Delta_{3}$ are given by
\begin{eqnarray} 2\Delta_{1,2}&=&G_d \Delta_{1,2}\int
\frac{d^3p}{(2\pi)^3}\biggl(\frac{8}{E^+(\overline{M},
\overline{\mu}_{1,2},\Delta_{1,2})} + \frac{8}{E^-(\overline{M},
\overline{\mu}_{1,2},\Delta_{1,2})}
+\nonumber\\&&\frac{8}{E^+(\overline{M}, \overline{\mu}_{1,2},2\Delta_{1,2})} +
\frac{8}{E^-(\overline{M}, \overline{\mu}_{1,2},2\Delta_{1,2})}\biggr) \,
, \label{delta12}
\end{eqnarray} and \begin{eqnarray}
\Delta_{3}= G_d \Delta_{3}\int \frac{d^3p}{(2\pi)^3}
\biggl(\frac{8}{E^+(\overline{M}, \overline{\mu}_{3},\Delta_{3})} +
\frac{8}{E^-(\overline{M}, \overline{\mu}_{3},\Delta_{3})}\biggr) \, ,
\label{delta3}
\end{eqnarray} respectively.

The magnitude of $\delta\rho$ used in Eqs.(\ref{murho12}) and (\ref{murho3})
could not be yielded from the above-given scheme itself.
It has been widely discussed in the literature that there exist
some less-symmetric BCS phases as possible candidates of
the second-densest ground state of quark matter. In the present work, we only consider the
CFL phase where the pseudo Goldstone excitations become condensed.
Since the $K^0$ mode is the strange-involved one ( as the usual $K^0$ meson ), a small $m_s$ triggers the
$K^0$-mode condensation in the CFL environment
( the so-called CFL$K^0$ phase ).
As usual, $K^0$ has the chemical potential
$\mu_K={m_s^2}/{2\mu}$ because of the chemical equilibrium.
Based on the chiral effective Lagrangian for Goldstone modes,
the hypercharge density from the $K^0$ condensation reads \cite{sch,kr,zhang04}
\begin{eqnarray}
\rho_{K}=f_\pi^2 \mu_K,
\label{rhok}
\end{eqnarray}
where the in-CFL-medium decay constant $f_\pi \simeq 0.21 \mu$ was obtained in the weak-coupling limit \cite{fpi}.
In the situation with realistic $m_{u,d,s}$ and medium $\mu$,
the condensation may be weaken and the value of $f_\pi$ might become no longer correct \cite{forbes,buba}.
Also, $K^0$ condensation is possible to be excluded because of the instanton effect \cite{ins}.
As long as CFL$K^0$ emerge, Eq.(\ref{rhok}) accounts for the change in the density of the strange-flavor quasi particles.
In the following calculations, we will identify Eq.(\ref{rhok}) with the density change
$\delta\rho$ introduced above.

Another simplification in the calculations involves the values
of $\overline{\mu}_{1,2}$ and $\overline{\mu}_{3}$. In principle,
the average chemical potentials need to be solved from
Eqs.(\ref{murho12}) and (\ref{murho3}) as well as the gap equations
self-consistently.
If ignoring the difference of ${\partial
\Omega_{1,2}}(\Delta_{1,2})/{\partial \overline{\mu}_{1,2}}$ from
the ideal result ${\partial \Omega_{1,2}(\Delta)}/{\partial \mu}$,
Eq.(\ref{murho12}) leads to an approximate relation
\begin{eqnarray}
\overline{\mu}_{1,2}\approx
{\mu}/{\left(1+\frac{\delta\rho}{\frac{\partial
\Omega_{1,2}(\Delta)}{\partial \mu}}\right)},\label{mu12}
\end{eqnarray}
in the strange-flavor-involved channel related with $\Delta_{1,2}$. Similarly, the value of
$\overline{\mu}_{3}$ is obtained by
\begin{eqnarray}
\overline{\mu}_{3}\approx {\mu}/{\left(1-\frac{\delta\rho}{\frac{\partial
\Omega_{3}(\Delta)}{\partial \mu}}\right)}, \label{mu3}
\end{eqnarray}
in the light-flavor channel related with $\Delta_{3}$. Because of the
negative values of ${\partial \Omega(\Delta)}/{\partial \mu}$, it is
obvious that $\overline{\mu}_{1,2}$ is larger than $\mu$ whereas
$\overline{\mu}_{3}$ is smaller than $\mu$.
This is a simple but nontrivial result which might be relevant for the onset of bound diquark states.

\subsection{The unlocking transition to 2SC }

As color-flavor unlocking takes place, it is usually believed that the so-called 2SC phase emerges.
The 2SC thermodynamic potential can be expressed analytically even though the current quark masses are introduced explicitly \cite{buba02}.
Different from the above description for CFL, the constituent quark masses $M_s$ and $M_{u,d}$ need to be introduced on the 2SC side.
Because only the light-flavor channel participates in pairing, the
dispersion relation for the paired quarks ( $rd-gu$ and $ru-gd$ ) has the similar form as
Eq.(\ref{e+-}). Correspondingly, the thermodynamic-potential contribution is written as
\begin{eqnarray}
\frac{\Delta_{2SC}^2}{4 G_d} + \Omega_3 ( M_{u,d},\mu, \Delta_{2SC} ),
\label{2sc1}
\end{eqnarray}
formally, where
$\Delta_{2SC}$ denotes the pairing gap on the 2SC side.
The unpaired quark species provide
the contributions
\begin{eqnarray}
- \sum \int \frac{d^3p}{(2\pi)^3}\left[ E^+( M_i,\mu,\Delta=0) + E^-(
M_i,\mu,\Delta=0)\right], \label{2sc2} \end{eqnarray} with $M_i=M_{u,d}$ (
for $bu$ and $bd$ ) and $M_i=M_s$ ( for $bs,rs$ and $gs$ ). Besides,
the mean-field contribution from the quark masses is
$2{M_{u,d}}^2/{8 G_s} + {(M_s-m_s)}^2/{8 G_s}$.

\begin{figure}[h]
\begin{center}
\includegraphics[angle=0, scale=0.6]{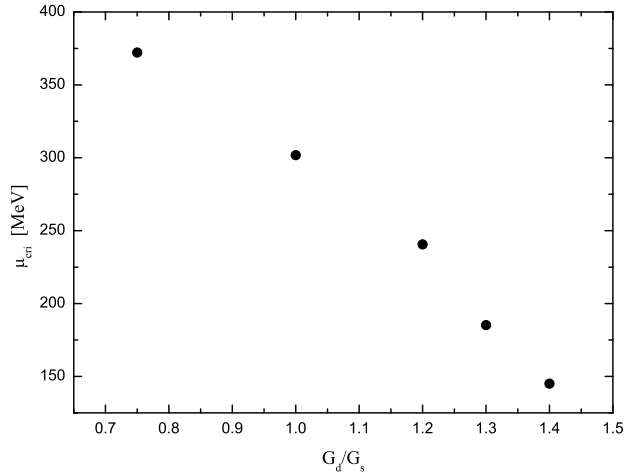}
\caption{Critical chemical potential
for the unlocking transition with different $G_d /G_s$. }
\end{center}
\end{figure}

Ignoring the 2SC electric/color neutrality and the $\beta$ equilibrium,
we simply treat color-flavor unlocking as a first-order phase
transition between CFL and 2SC. The critical chemical potential
$\mu_{cri}$ for the unlocking transition may be obtained from the
Gibbs' s pressure equilibrium. Now we adopt the model parameters
used in Ref.\cite{kita08}, i. e. the momentum cutoff
$\Lambda=600$MeV and the quark-antiquark coupling constant
$G_s=6.4$GeV$^{-2}$ and treat the diquark coupling " constant "
$G_d$ as a free parameter. Based on the above-defined CFL and 2SC
phase, the numerical calculation shows that the coupling-constant
ratio $G_d/G_s$ has an obvious influence on the locations of
$\mu_{cri}$. With a commonly used value in literatures like
$G_d/G_s= 0.75$, the unlocking transition is found to occur at
$\mu_{cri} \sim 400$MeV. As the coupling-constant ratio increases,
the critical value is found to be strongly suppressed. When the
value of $G_d/G_s$ becomes larger than $1.0$, the solution of
$\mu_{cri}$ is shown to be unrealistically small ( see Fig.
1, where the current mass $m_s=50$MeV is considered ). To understand
this result, it is worthy being noticed that CFL is calculated in a
different way from the literature. Due to the ( almost ) same
dispersion behaviors
for all the paired quarks, it is natural that the CFL existence is welcome when a stronger attractive interaction between quarks is
introduced.

To this end, one may doubt the validity of $\mu_{cri} < 300$MeV given in
Fig. 1. For such a small chemical potential, according to the
current knowledge of QCD phase diagram, there seems to exist
the confined hadronic phase rather than the quark
color superconducting phase. Keeping
it in mind that our concerned three-flavor superconductor is a
fictional phase being valid near the flavor $SU(3)$  limit, the
magnitude of $\mu$ could not be identified with the baryon density
of strongly interacting matter in realistic situations. In addition,
strong coupling strength like $G_d/G_s \sim 1.1-1.5$ had
been used for studying quark superconducting phase and the BEC
possibility in it \cite{abuki02,itak03,zhuang07,kita08}. There, exploring
the BCS-BEC crossover physics was actually limited in
the scope of strong-coupling quark superconductivity and the phase
transition to the confined hadronic phase was not incorporated yet.
As a theoretical interest, we will examine the BCS-BEC crossover physics in the strong-coupling
quark superconductor ( including CFL and 2SC ) all through the present work.

\section{\bf  Application to the BCS-BEC physics }

According to the energy of quark quasiparticle excitations
$E^{+}$ given in Eqs.(\ref{e+-3}) and (\ref{e+-12}), the minimum of the
dispersion relation is located at zero momentum when the average mass
$\overline{M}$ is larger than the average chemical potential
$\overline{\mu}$.
This indicates that Cooper
pairing is no longer restricted to a vicinity of the Fermi surface.
For the relativistic fermion gas, thus, the necessary condition of onsets of bound
diquark states and their BEC is written as
\begin{eqnarray} \overline{M}> \overline{\mu},\ \label{bec} \end{eqnarray}
Also, Eq.(\ref{bec}) might be understood from the viewpoint that diquark
states decay into a pair of quarks.
The threshold energy for this decay process is expressed as
$2(\overline{M}-\overline{\mu})$ since the excitation energy of a
quark is $M-\mu$ at zero momentum. To ensure stability of the
diquark system, the threshold energy is required to be positive.

\begin{figure}[!ht]
\begin{center}
\includegraphics[angle=0, scale=0.6]{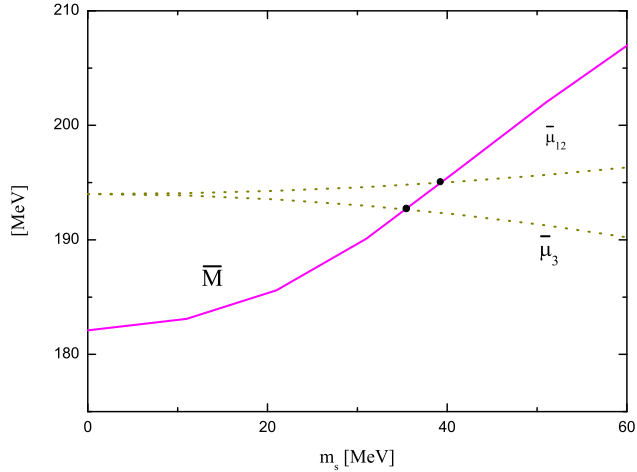}
\caption{Responses of average quark mass ( solid line ) and average
chemical potentials ( dot lines ) to $m_s$ in the
CFL phase with $G_d /G_s = 1.3$ and $\mu \simeq 195$MeV. }
\end{center}
\end{figure}

In the present description of CFL, the average masses have been assumed to have a
common value.
At the same time, the effect of nonzero $m_s$ ( exactly, $m_s^2/2\mu$ ) has been attributed to the difference of the average
chemical potentials.
At a given $\mu$, the diquark behaviors are shown in Fig. 2. There, the strange current mass
has been treated as a " free " parameter.
For the two kinds of diquark states, the average chemical potentials are found to be separated. It is observed that
$\overline{M}>\overline{\mu}_{3}$ is relatively easy to be satisfied.
This result is not trivial if comparing it with the popular one based on the ordinary CFL description. In the literature, $m_s$ was introduced explicitly and the constituent masses $M_{s}$ and $M_{u,d}$ took the different values. Noticing that the average mass $(M_{s}+M_{u,d})/2$ is usually larger than $M_{u,d}$, Eq.(\ref{bec}) would be easier to be satisfied in the strange flavor involved channel related with $\Delta_{1,2}$. In the present work, however, it is the light-flavor channel related with $\Delta_{3}$ to form the BEC-like bound diquark states firstly.
Because the BCS pairs in the strange-involved channel comes to be unfavored with raising $m_s$, the tendency shown in Fig. 2 is quite reasonable. It holds valid in the case of decreasing $\mu$, as will be seen in Fig. 3.

\begin{figure}[!ht]
\begin{center}
\includegraphics[angle=0, scale=0.6]{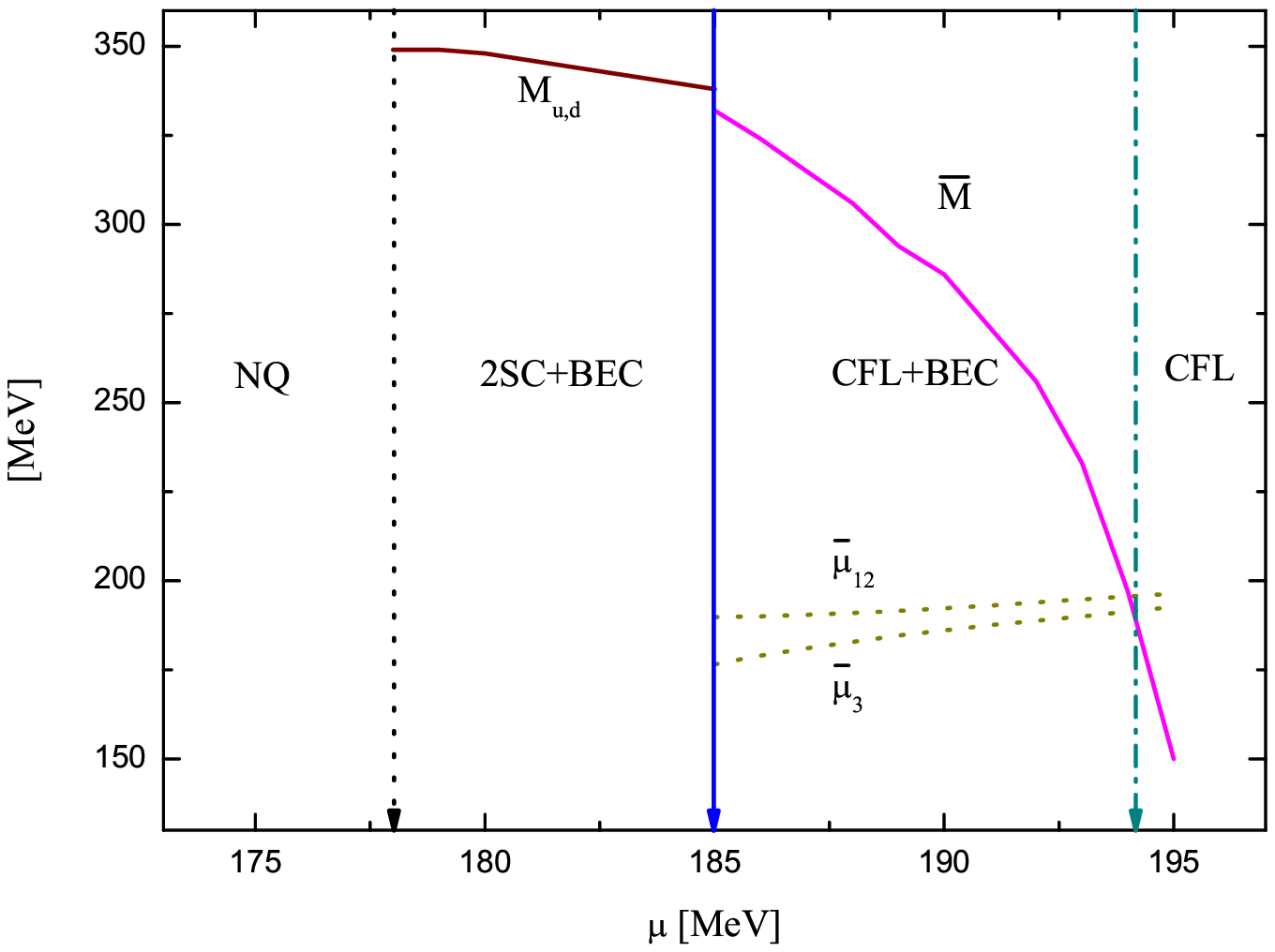}

\includegraphics[angle=0, scale=0.6]{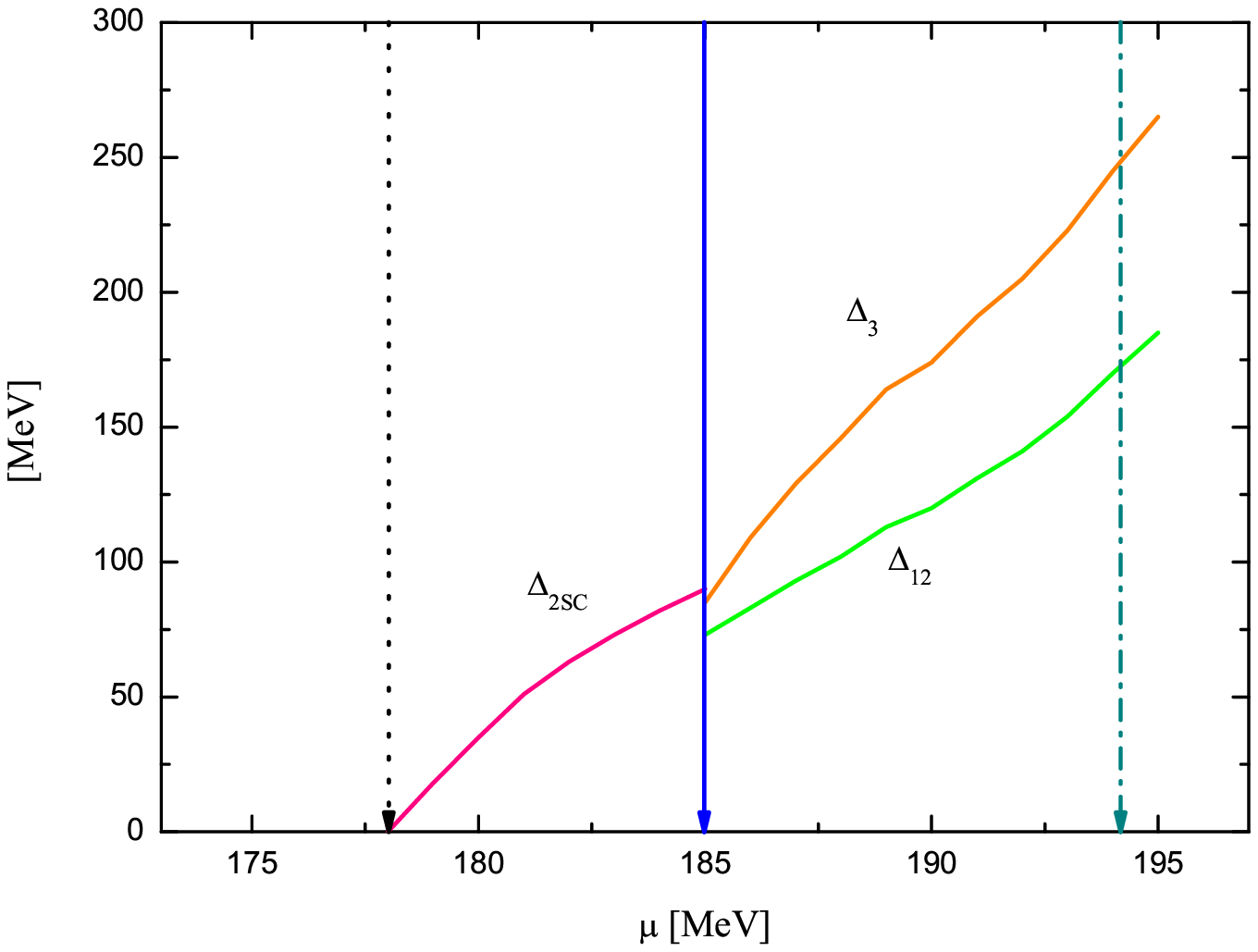}
\caption{Behaviors of average quark mass and average chemical potentials ( upper panel ) as well as gaps ( lower ) in strong-coupling quark superconductor ( $G_d /G_s = 1.3$ ). The dotted-dashed (dark cyan) line denotes the BCS-BEC crossover occurred in the CFL environment while
the solid (blue) and dotted (black) lines correspond to the CFL-2SC transition and the transition to normal quark matter (NQ) respectively.}
\end{center}
\end{figure}

Now we turn to explore the BEC possibility by taking color-flavor unlocking into account.
Based on Fig. 1 , the critical value $\mu_{cri}$ is small enough as long as a large coupling-constant ratio is considered. In this case, we should  investigate the BCS-BEC crossover in the environment where the CFL phase is energetically favorable. In Fig. 3,
the average quark mass as well as the average chemical potentials
are given in the situation with $m_s=50$MeV.
With a large ratio $G_{d}/G_{d}=1.3$, the crossover point $\mu_{X}^{CFL}$ defined by
$\overline{M}=\overline{\mu}_3$ is found to be larger than the critical value $\mu_{cri} \simeq 185$MeV.
In the narrow region of $\mu_{cri}<\mu < \mu_{X}^{CFL}$, there exists Bose-Einstein condensation of bound diquark states. This is referred as CFL+BEC in Fig. 3.
Also, the gap
$\Delta_3$ in the light-flavor channel is found to have the similar
order as $\mu$ ( the lower panel of Fig. 3 ).
In this sense, we conclude that ( at least ) the light-flavor diquark states
undergo Bose-Einstein condensation in the CFL environment.
On the other hand, we must examine whether or not a BCS-BEC crossover occur in the unlocked environment.
In the 2SC phase, the average quark mass is
$M_{u,d}$ ( in the light-flavor channel ) while the average chemical potential takes the value of $\mu$.
From the competition of $M_{u,d}$ and ${\mu}$, the resulting BCS-BEC crossover is located at
$\mu_{X}^{2SC} \simeq 230$MeV( which is not given in Fig. 3 ). Because the 2SC phase
does not exist for $\mu >\mu_{cri}$, such a crossover point is unphysical.
In the whole region where 2SC is favored, nevertheless, BEC of light-flavor diquark states still makes sense since $M_{u,d}>\mu$ is always satisfied there.
To distinguish the above-discussed CFL+BEC phase, the Bose-Einstein condensed phase is shown
as the 2SC+BEC phase in Fig. 3. Also, it is noticed that 2SC+BEC becomes terminated at $\mu \simeq 178$MeV where the gap $\Delta_{2SC}$ comes to be zero.


\begin{figure}[!ht]
\begin{center}
\includegraphics[angle=0, scale=0.6]{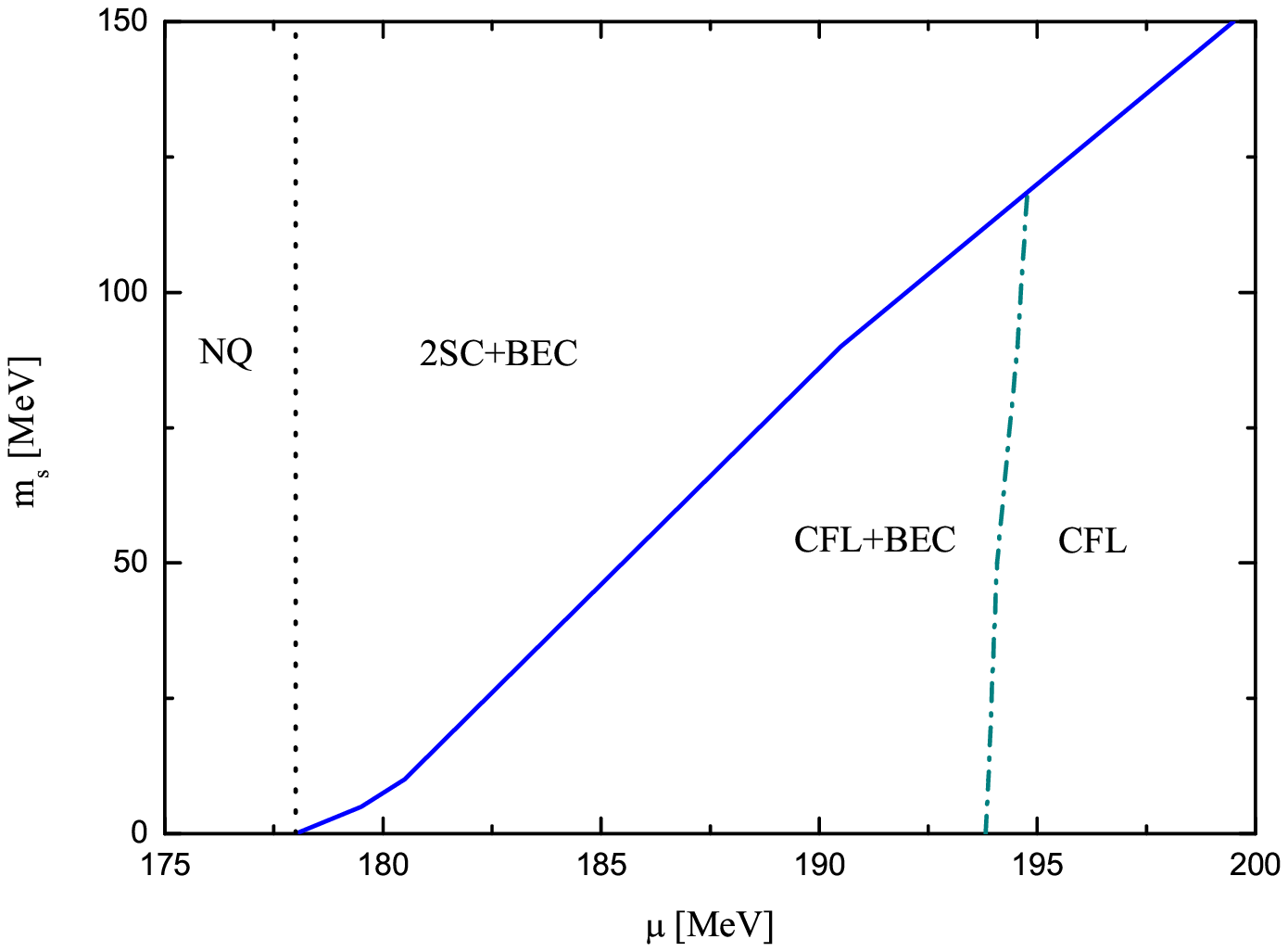}
\caption{Predictive phase diagram as a direct expansion of Fig. 3 . }
\end{center}
\end{figure}

For demonstrative purpose, then, we give a predictive phase diagram in the $(\mu, m_s)$ plane.
Starting from the ideal situation ( $m_s=0$ ) and raising $m_s$ gradually, on the one hand,
the light-flavor diquark states may undergo Bose-Einstein condensation in the CFL environment.
As shown by the dotted-dashed line in Fig. 4 ,
the CFL-CFL+BEC crossover is realized for ( slightly ) larger $\mu$.
On the other hand, varying $m_s$ influences the location of $\mu_{cri}$ more obviously. As shown by the solid line in Fig. 4 ,
the unlocking transition takes place at a larger $\mu_{cri}$ when $m_s$ increases.
By considering the two effects, it is observed that the CFL+BEC regime becomes reduced for medium $m_s$.
Once the current mass is chosen to be larger, the CFL+BEC phase is actually impossible to be form.
For the realistic parameter $m_s=120-150$MeV, $\mu_{cri} < \mu_{X}^{CFL}$ is no longer satisfied so that
the crossover from CFL to CFL+BEC does not occur.
As for the 2SC+BEC existence, it is noticed that the location of 2SC-2SC+BEC crossover is not influenced by the strange current mass directly.
As above mentioned, the value of $\mu_{X}^{2SC}$ defined by $M_{u,d}=\mu$ is about $230$MeV. It is larger than the critical value $\mu_{cri}$ even for
realistically large $m_s$. As shown in Fig. 4, therefore, 2SC+BEC emerges as long as
color-flavor unlocking has happened and it behaves as the only Bose-Einstein condensed phase in the situation with realistic
value of $m_s$.
Such kind of phase holds valid until a transition to the normal phase takes place ( the dot line ).

In fact, the crossover from the CFL-type BCS phase to the CFL+BEC phase is not a surprise.
Within the present framework,
color-flavor unlocking and Bose-Einstein condensation have been treated as the two of
independent phenomena. The former corresponds to the BCS breaking in the strange-flavor-involved
channel while the latter has been actually limited in the light-flavor channel.
As long as $\mu_{cri} < \mu_{X}^{CFL}$ is satisfied, the CFL+BEC possibility arises, namely
( at least ) the light-flavor diquark states undergo Bose-Einstein condensation in the CFL environment.
Also, the present result may be understood from the viewpoint that the structure change of Copper pairs determines onset of a BCS-BEC crossover \cite{itak03}.
For our concerned CFL and 2SC superconductors, the dispersion relations for diquarks are expressed analytically
so that the light-flavor quasiparticle energies $E^+(p)$ have the similar forms.
Also, the pairing gap functions $\Delta_3(p)$ ( for CFL ) and $\Delta_{2SC}(p)$ ( for 2SC ) have the similar structures in the momentum space
( because the identified matrix elements of quark-quark interaction are employed ).
As the consequence, there should be not essential difference between the two of light-flavor Cooper-pair wave functions
defined by $\Delta_3(p)/2E^+(p)$ ( for CFL ) and $\Delta_{2SC}(p)/2E^+(p)$ ( for 2SC ).
Similar as that predicted in the unlocked environment, therefore, it is very natural to suppose that the spatial structure of Copper-pair wave function evolves continuously from the CFL phase to the CFL+BEC phase. As a whole, the Bose-Einstein condensed phases exist
in the environment where color-flavor-locked BCS pairing pattern firstly remains and then it becomes broken.
In other words, CFL+BEC is regarded as an intergrade phase before 2SC+BEC exists eventually. This is just the picture given by Figs. 3 and 4.

\begin{figure}[!ht]
\begin{center}
\includegraphics[angle=0, scale=0.6]{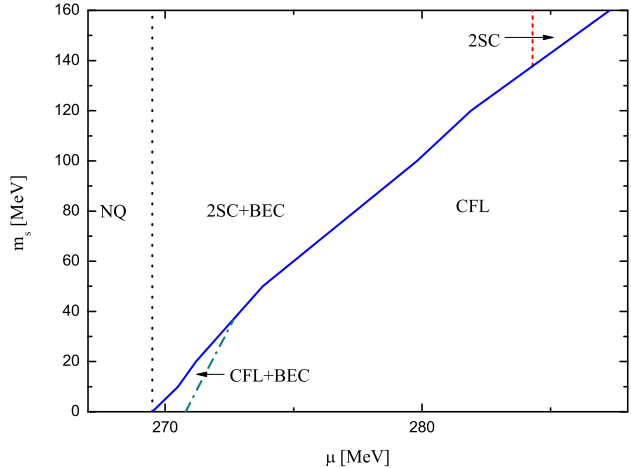}
\caption{Similar as Fig. 4, but with $G_d/G_s = 1.1$. The short dashed (red) line denotes the crossover from 2SC to 2SC+BEC.}
\end{center}
\end{figure}

Finally, we adopt the coupling-constant ratio $G_d/G_s=1.1$ to investigate other possibilities of BCS-BEC crossover.
Besides the CFL-CFL+BEC crossover occurred for very small $m_s$, a BCS-BEC crossover becomes possible in the unlocked environment.
With such a parameter $G_{d}/G_{d}$, the solutions of $\mu_{cri}$ and $\mu_{X}^{2SC}$ become comparable for large $m_s$. When $m_s$ is realistically large, as shown in Fig. 5, it is not CFL+BEC but 2SC to behave as an intergrade phase.
Actually this result is same as the previous one obtained in Ref.\cite{kita08}, where CFL was calculated nonanalytically in the situation with the realistic current masses. Once the smaller ratios are chosen, the solutions of $\mu_{cri}$ become large systematically ( see Fig. 1 ). Thus,  $\mu_{cri} < \mu_{X}^{CFL}$ is usually not satisfied and CFL+BEC does no longer emerge.
Within the NJL framework, it has been noticed that Bose-Einstein condensation of diquarks occurs only if
the coupling constant $G_d$ is sufficiently large \cite{zhuang07,abuki07,kita08}.

\section{\bf  Summary and discussion }

Within a three-flavor NJL model, an approximate description for the color-flavor-locked-type BCS phase has been proposed near the flavor $SU(3)$  limit ( near the chiral limit ). Assuming that analytical dispersion relations are valid for all the diquark species,
the common constituent quark mass $\overline{M}$ is used while the physical influence from a small $m_s$ is attributed
to difference of the average chemical potentials $\overline{\mu}_{1,2}$ and $\overline{\mu}_{3}$.
At mean-field level, the existence of light-flavor diquark BEC phase becomes inevitable as long as the attractive interaction between quarks is strong enough.
In particular, a possibility was reported that
light-flavor diquark BEC phase emerges in the environment where the color-flavor-locked pairing pattern remains.
It is a consequence of generalizing three-flavor superconductor to the case of arbitrary small values of $m_s$ and/or relatively low values of $\mu$.
In spite of that CFL+BEC is limited in a very narrow regime, this result
enriches the phase structure of strong-interacting matter and might be helpful for better understanding
quark superconductor in the theoretical sense.

If taking into account other effects such as the interplay between chiral and diquark condensates,
it may be expected that existence of light-flavor BEC phase is enhanced  more or less. In Ref.\cite{abuki10},
H. Abuki, G. Baym, T. Hatsuda and N. Yamamoto introduced the coupling between chiral and diquark condensates
through the axial anomaly and found that the BEC regime is enlarged.
While finishing our paper, we learned that a more recent study was done by H. Basler and M. Buballa \cite{babu10}. There, the
possibility of CFL+BEC ( called as CFL$_{BEC}$ ) was reported in a three-flavor NJL model with the axial anomaly.
Due to the anomaly effects, also, they found that the 2SC pairing could be still favored even for small current masses.
If so, it is plausible that CFL+BEC always behave as an intergrade
phase which emerges before 2SC+BEC.
Of course, comparison with the results of Refs.\cite{abuki10,babu10} requires going beyond the scope of the present work. This is worth being further examined.

Finally, we stress again that the present description for three-flavor superconductor makes sense
only when the flavor $SU(3)$ symmetry holds approximately.
To make the results more reliable, it is important to take into account effects of color/electric chemical potentials, color/electric neutralities of quark matter as well as possible gapless phenomenon.
In this case, the physical influences caused by explicitly chiral breaking can not be simply attributed to a difference of the light- and strange-flavor densities. Instead, one should derive the average chemical potentials $\overline{\mu}$ from the Fermi-momentum mismatches induced by all the effects.
In the realistic situations, there may exist variants of two-flavor superconductor ( like the 2SCsu phase with only $\Delta_2$ nonzero
and the 2SCds phase with only $\Delta_1$ ) and non-BCS superconductor. In this case,
color-flavor unlocking should be no longer realized by an ordinary CFL-2SC transition.
In the further investigation, a more systematic analysis on the BEC occurrence is needed in view of
the above considerations.

\vspace{0.5cm} \noindent {\bf Acknowledgements} \vspace{0.5cm}

The authors thank Prof. Xue-Qian Li for useful discussions.

\end{document}